\documentclass[a4paper,twocolumn]{esapub} % European paper
\usepackage{natbib}
\usepackage{epsfig}

\title{SCIENCE PROSPECTS FOR SPI}
\author{J\"urgen Kn\"odlseder}
\author{Gilbert Vedrenne}
\affil{Centre d'Etude Spatiale des Rayonnements, B.P. 4346, 31028 
Toulouse Cedex 4, FRANCE}

\newcommand{\al}{\mbox{$^{26}$\hspace{-0.2em}Al}}
\newcommand{\fe}{\mbox{$^{60}$Fe}}
\newcommand{\ti}{\mbox{$^{44}$Ti}}
\newcommand{\coa}{\mbox{$^{56}$Co}}
\newcommand{\cob}{\mbox{$^{57}$Co}}
\newcommand{\be}{\mbox{$^{7}$Be}}
\newcommand{\na}{\mbox{$^{22}$Na}}
\newcommand{\nuc}[2]{\mbox{$^{#1}$#2}}
\newcommand{\Msol}{\mbox{$M_{\sun}$}}

\newcommand{\pcmq}{\mbox{cm$^{-2}$}}

\newcommand{\psec}{\mbox{s$^{-1}$}}

\newcommand{\funit}{\mbox{ph \pcmq \psec}}
 
\newcommand{\frad}{\mbox{ph \pcmq \psec rad$^{-1}$}}
\def\MeV{\mbox{Me\hspace{-0.1em}V}}

\def\deg{\ensuremath{^\circ}}
\def\sun{\hbox{$\odot$}}

\begin{document}


\maketitle

%%%%%%%%%%%%%%%%%%%%%%%%%%%%%%%%%%%%%%%%%%%%%%%%%%%%%%%%%%%%%%%%%%%%%%%%%%%%%%%%
\begin{abstract}
After the recent beautiful results on gamma-ray lines obtained with 
{\em Compton GRO} (OSSE and COMPTEL), the {\em INTEGRAL} mission with the 
imaging-spectrometer SPI will set the next milestone, combining improved 
sensitivity and angular resolution with a considerable increase in 
spectral resolution. 
SPI is expected to provide significant new information on galactic 
nucleosynthesis processes and star formation activity, as traced by 
the distributions of annihilation radiation and radioactive isotopes 
such as \al\ and \fe. 
The unprecedented spectral resolution will allow the study of dynamic 
processes in stellar mass ejections and will provide access to kinematic 
distance estimates for gamma-ray line sources.

The study of supernovae and their remnants will be prime objectives for 
SPI observations. 
Nearby type Ia SN, within 15 Mpc or so, are in reach of the instrument 
and a few such events are expected during the lifetime of {\em INTEGRAL}. 
Young galactic supernova remnants, possibly hidden by interstellar dust, 
may be unveiled by their characteristic gamma-ray line signature from 
the radioactive decay of \ti, as has been demonstrated by COMPTEL for 
Cas A and possibly RX J0852.0-4622. 
Even SN1987A, which has already faded in the short lived radioisotopes 
\coa\ and \cob, could again become a target of interest for gamma-ray 
line astronomy, since the longer lived isotope \ti\ may be in reach of 
SPI.

Classical novae are also among the SPI targets, which may observe the 
gamma-ray lines from radioactive \be\ and \na. 
Such observations can constrain the physics of the nova explosions 
and will allow to evaluate their role as nucleosynthesis sites.
The interaction of cosmic rays with the dense matter in molecular 
clouds may be another source of gamma-ray lines that is potentially 
accessible to SPI. 
Finally after the SIGMA results on Nova Muscae and 1E1740.7-2942, and 
a possible 2.223 \MeV\ line detection by COMPTEL, the search for lines 
from X novae is another way to participate in the understanding of 
the physical conditions in these close binary systems.
\end{abstract}

%%%%%%%%%%%%%%%%%%%%%%%%%%%%%%%%%%%%%%%%%%%%%%%%%%%%%%%%%%%%%%%%%%%%%%%%%%%%%%%%
\section{Design considerations for SPI}

The successful {\em Compton Gamma-Ray Observatory} (CGRO) provided a 
brilliant demonstration of the great scientific potential promised by 
the field of gamma-ray line astronomy.
The COMPTEL telescope established the first all-sky map in the light 
of the \al\ decay line at 1.809 \MeV, revealing diffuse structured
emission from the entire galactic plane arising from massive star 
nucleosynthesis (e.g.~Pl\"uschke et al., these proceedings).
The detection of the 1.157 \MeV\ line towards Cas A \citep{iyudin94} 
was the first observation of radioactive \nuc{44}{Ti} decay 
related to a young supernova remnant, and the following possible 
detection of 1.157 \MeV\ line emission towards Vela \citep{iyudin98} 
triggered for the first time the discovery of a new young supernova 
remnant (RX J0852.0-4622).
OSSE provided the first maps in the 511 keV line and positronium 
continuum emissions, identifying at least two galactic emission 
components (extended bulge and disk) with indications for a possible
third component situated above the galactic plane (see Milne et al., 
these proceedings).
Although OSSE observed SN~1987A more than 4 years after explosion, it 
still was capable to detect the \nuc{57}{Co} decay line at 122 keV, 
leading together with the \nuc{56}{Co} observations by {\em SMM} to 
the first determination of an isotopic abundance ratio through 
gamma-ray line measurements.

All these discoveries confirmed that gamma-ray line observations are 
a powerful probe of nuclear astrophysics, and the experience gained 
from the observations sets the design guideline for the spectrometer 
SPI on {\em INTEGRAL}.
First, the most prominent gamma-ray lines (1.809 \MeV\ and 511 keV) 
turned out to be extended over angular scales of $\sim10$ degrees,
hence an instrument was needed that provides good sensitivity to 
extended emission features.
Second, gamma-ray line fluxes showed to be quite low -- a few $10^{-5}$ 
\funit\ at maximum -- requiring an instrument that is sensitive to 
fluxes well below that level.
And third, emission regions are often quite confused, asking for 
sufficient angular resolution to spatially resolve the emission,
allowing for an adequate identification of candidate sources.
Additionally, the {\em CGRO} instruments did not provide sufficient 
spectral resolution to resolve gamma-ray line profiles and to 
determine line centroids, hence an instrument was needed that can also 
access this extremely valuable complementary information.

Having these considerations in mind, the following design was chosen 
for SPI:
\begin{itemize}
\item a coded mask to modulate the gamma-ray emission, allowing for
      good angular resolution ($2\deg$) and source localisation 
      throughout the field-of-view,
\item a wide fully coded field-of-view of $16$ degrees ($34\deg$ 
      partially coded), allowing for imaging of extended emission 
      regions,
\item the use of cooled Germanium detectors for good gamma-ray line 
      sensitivity by effectively reducing the number of background 
      counts thanks to an excellent energy resolution (2 keV at 1 \MeV),
      which also allows studies of line shifts and line profiles.
\end{itemize}
It has to be emphasised that the sensitivity to detect gamma-ray lines 
is determined by the instrumental background underlying the line.
Based on Monte-Carlo simulations of the instrumental background, the 
narrow-line sensitivity of SPI has been estimated to better than 
$8\,10^{-6}$ \funit\ above 200 keV \citep{jean96}.
Since a broadened line includes more background counts than a narrow 
line, the sensitivity of SPI is effectively reduced for lines that are
broader than the detector spectral resolution.
As a rule of thumb one can estimate the sensitivity to broad lines
(expressed as the minimum detectable flux $\Phi_{\mbox{broad}}$)
using
\begin{equation}
 \Phi_{\mbox{broad}} = \Phi_{\mbox{narrow}} 
 \sqrt{\Delta E_{\mbox{broad}} / \Delta E_{\mbox{GeD}}} ,
 \label{eq:broad}
\end{equation}
where 
$\Phi_{\mbox{narrow}}$ is SPI's narrow line sensitivity,
$\Delta E_{\mbox{broad}}$ is the line width, and
$\Delta E_{\mbox{GeD}}$ is the instrumental line width
(this formula is of course only valid for 
$\Delta E_{\mbox{broad}} \ge \Delta E_{\mbox{GeD}}$).
For a more detailed description of the expected SPI performances, the 
reader is referred to \citet{mandrou97}.

%%%%%%%%%%%%%%%%%%%%%%%%%%%%%%%%%%%%%%%%%%%%%%%%%%%%%%%%%%%%%%%%%%%%%%%%%%%%%%%%
\section{Radioactivities}

%%%%%%%%%%%%%%%%%%%%%%%%%%%%%%%%%%%%%%%%%%%%%%%%%%%%%%%%%%%%%%%%%%%%%%%%%%%%%%%%
\subsection{The gamma-ray line menu}

%%% Table: Gamma-ray lines from radioactivities %%%%%%%%%%%%%%%%%%%%%%%%%%%%%%
\begin{table*}
  \begin{center}
    \caption{The menu of gamma-ray lines from radioactivities that may be 
             accessible to gamma-ray astronomy (ordered by ascending lifetime).
             Theoretical nucleosynthesis yield estimates are quoted for 
             different source types; the yields for AGB stars are 
             split into low-mass AGBs ($< 5 \Msol$, left column) and 
             high-mass AGBs ($> 5 \Msol$, right column).
             Positron emitters are marked by $\dagger$.}\vspace{1em}
    \renewcommand{\arraystretch}{1.2}
    \begin{tabular}{l l l l l l l l l l}
      \hline 
      Isotope & Lifetime $\tau$ & Lines (keV) & \multicolumn{7}{c}{Typical 
      yields (\Msol)} \\
      \cline{4-10}
      & & & \multicolumn{2}{c}{AGB} & 
      WR & SN Ia & SN Ib/c & SN II & Nova \\
      \hline
      \nuc{57}{Ni} & 2.14 d & 1378 & & & 
        & 0.02 & $5\,10^{-3}$ & $5\,10^{-3}$ & \\	
      \nuc{56}{Ni} & 8.5 d  &	158, 812 & & &
        & 0.5 & 0.1 & 0.1 & \\	
      \nuc{59}{Fe} & 64.2 d & 1099, 1292 & & &
        & & $5\,10^{-5}$ & $5\,10^{-5}$ & \\	
      \nuc{7}{Be}  & 77 d   & 478 & & &
        & & $10^{-7}$ & $5\,10^{-7}$ & $5\,10^{-11}$ \\	
      \nuc{56}{Co}$^{\dagger}$ & 112 d & 847, 1238 & & &
        & 0.5 & 0.1 & 0.1 & \\	
      \nuc{57}{Co} & 392 d & 122 & & &
        & 0.02 & $5\,10^{-3}$ & $5\,10^{-3}$ & \\	
      \nuc{22}{Na}$^{\dagger}$ & 3.76 yr & 1275 & & &
        & $10^{-8}$ & $10^{-6}$ & $10^{-6}$ & $5\,10^{-9}$ \\	
      \nuc{60}{Co} & 7.61 yr & 1173, 1332 & & &
        & & $10^{-5}$ & $10^{-5}$ & \\				
      \nuc{44}{Ti}$^{\dagger}$ & 87 yr & 1157 & & &
        & $10^{-5}$ & $5\,10^{-5}$ & $5\,10^{-5}$ & \\	
      \nuc{26}{Al}$^{\dagger}$ & $10^6$ yr & 1809 & $10^{-7}$ & 
        $4\,10^{-6}$ &
        $10^{-4}$ & & $5\,10^{-5}$ & $5\,10^{-5}$ & $10^{-8}$ \\	
      \nuc{60}{Fe} & $2.2\,10^6$ yr & 1173, 1332 & & & $10^{-10}$
        & $5\,10^{-3}$ & $5\,10^{-5}$ & $5\,10^{-5}$ & \\	
      \hline \\
    \end{tabular}
    \label{tab:radioactivities} 
  \end{center}
\end{table*}
%%%%%%%%%%%%%%%%%%%%%%%%%%%%%%%%%%%%%%%%%%%%%%%%%%%%%%%%%%%%%%%%%%%%%%%%%%%%%%

From the roughly 2500 isotopes that are known today, only about $1/10$ 
are stable while all others decay radioactively or undergo spontaneous 
fission.
Many of the decays are accompanied by gamma-ray line emission due to 
nuclear de-excitations of the daughter nuclei, but only few isotopes
are indeed accessible to gamma-ray astronomy.
In fact, there are three basic conditions that have to be fulfilled 
for the detectibility of a radioisotope.
First, a hot and dense medium with sufficiently low entropy is required to 
allow for the synthesis of fresh radioisotopes.
Such a medium can be found in stellar interiors, at the base of the 
accreted envelope of white dwarfs in close binary systems, or even in 
accretion disks around compact objects.
The nuclear reaction networks in operation are characteristic for the 
composition, density, and temperature at the burning site, hence 
the observation of isotopic abundance patterns provide direct insight 
into the nucleosynthesis conditions.
Second, the fresh radioisotopes have to be removed quickly from the 
formation site to prevent destruction by nuclear reactions or natural 
decay.
This generally implies convection followed by mass ejection, either 
in form of stellar winds or explosions, and requires lifetimes of at 
least several days, better several months.
Additionally, nucleosynthesis sites are generally optically thick to 
gamma-rays, hence escape of the radioisotopes to optically thin 
regions is mandatory for gamma-ray line observations.
Consequently, radioisotopes can probe stellar convection and ejection 
processes, providing important information about the involved stellar 
physics.
Third, the lifetime has to be short enough and the abundance of the 
isotope has to be high enough to assure a sufficient radioactive 
decay activity that is in reach of modern gamma-ray telescopes.

These constraints result in a list of candidate isotopes that may 
actually be accessible to gamma-ray line astronomy (cf. Table
\ref{tab:radioactivities}).
The observation of radioactivity lines implies various time-scales.
For lifetimes that are short compared to the event frequency
(\nuc{57}{Ni} - \nuc{57}{Co}), individual transient gamma-ray line 
sources are expected, mainly in form of supernovae or novae.
The observation of gamma-ray line lightcurves provides then an important 
complementary information for studying the physics of these explosive 
events (e.g. Leising \& Share, 1990).
For lifetimes of the order of the event frequency
(\nuc{22}{Na} - \nuc{44}{Ti}) several rather steady (over the lifetime 
of a typical gamma-ray mission) gamma-ray line sources are expected in 
form of supernova remnants or recent nova events.
For lifetimes that are long compared to the event frequency 
(\nuc{26}{Al} - \nuc{60}{Fe}) the superposition of numerous individual 
sources will lead to a diffuse glow of gamma-ray line 
emission along the galactic plane.
Additionally, the longlived radioisotopes may travel considerable 
distances away from the production sites before they decay 
($\sim10-100$ pc), leading to intrinsically extended sources.

Some of the most prominent radioactive isotopes 
(\nuc{56}{Co}, \nuc{22}{Na}, \nuc{44}{Ti}, and \al)
are also positron emitters, and the annihilation of the positrons with 
electrons in the interstellar medium (ISM) may lead to 511 keV line 
emission, accompanied by a positronium continuum emission below 511 keV 
due to 3 photon decays.
The typical lifetime for positrons in the ISM is of the order of 
$10^5$ yr \citep{guessoum91}, hence the annihilation radiation 
provides a long-lasting reverberation of the short-lived 
radioisotopes.
In particular, since the positron lifetime considerably exceeds
the period of supernova or nova events, numerous sources should 
contribute to the annihilation radiation, leading to a diffuse glow of 
extended emission.

Note that event frequencies depend of course on the volume that is 
accessible to the gamma-ray telescope, making the above 
considerations mainly valid for galactic nucleosynthesis events.
If the accessible volume is extended to the entire Universe, diffuse 
gamma-ray emission is even expected from such short-lived 
radioisotopes as \nuc{56}{Co}.
In this case, however, the cosmological redshift will smear-out the 
lines, leading to continuum emission that could well explain the 
cosmic gamma-ray background radiation in the 100 keV - 1 \MeV\ energy 
range (Watanabe et al., 1999; Lichti et al., these proceedings).

%%%%%%%%%%%%%%%%%%%%%%%%%%%%%%%%%%%%%%%%%%%%%%%%%%%%%%%%%%%%%%%%%%%%%%%%%%%%%%%%
\subsection{\al\ -- studying the recent galactic star-formation 
history}
\label{sec:al}

%%% Figure: 26Al prospectives %%%%%%%%%%%%%%%%%%%%%%%%%%%%%%%%%%%%%%%%%%%%%%%%%%
\begin{figure*}
  \epsfxsize=8cm \epsfclipon
  \epsfbox{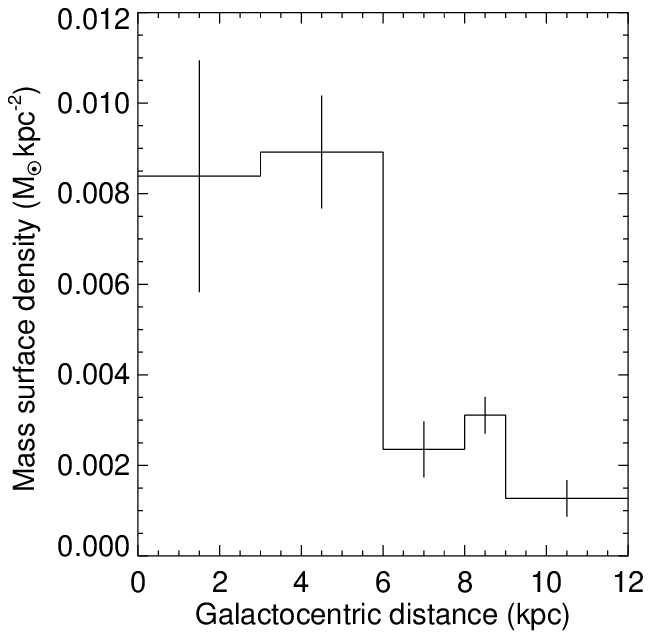}
  \hfill
  \epsfxsize=8cm \epsfclipon
  \epsfbox{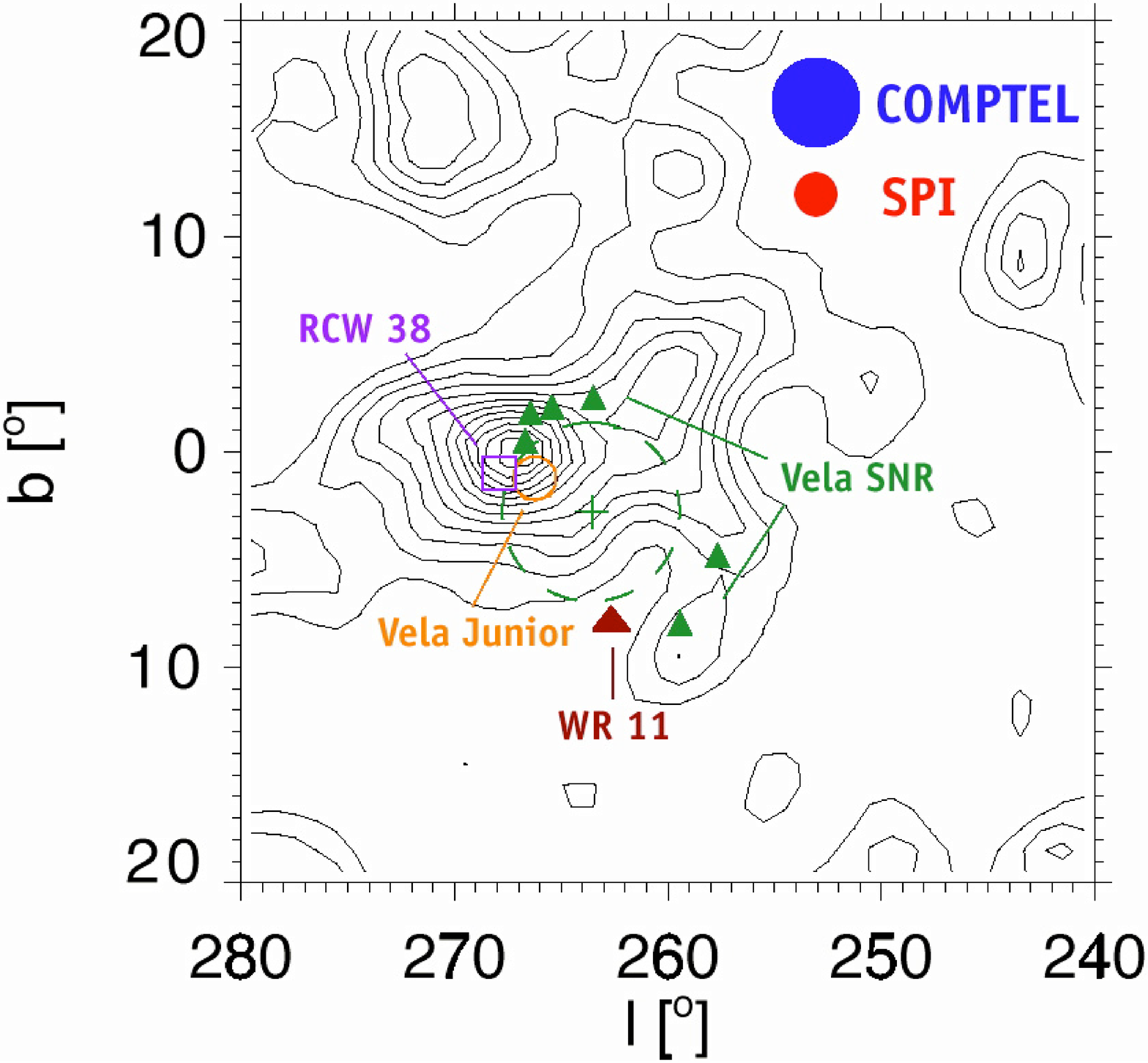}
  \caption{\label{fig:al26}
    SPI prospects for 1.809 \MeV. 
    The left panel shows the radial \al\ density profile as derived 
    from COMPTEL data \citep{knoedl97}.
    With improved sensitivity, SPI will refine this distribution, 
    allowing for a detailed study of the galactic star formation 
    activity.
    The right panel shows a COMPTEL image of the 1.809 \MeV\ emission 
    in the Vela region with potential \al\ sources superimposed 
    \citep{oberlack97}.
    The better angular resolution and sensitivity of SPI will allow to 
    identify counterparts of 1.809 \MeV\ emission, constraining 
    nucleosynthetic yields for individual objects or specific groups 
    such as OB associations and young open clusters (the filled 
    circles illustrate the SPI and COMPTEL angular resolutions).
  }
\end{figure*}
%%%%%%%%%%%%%%%%%%%%%%%%%%%%%%%%%%%%%%%%%%%%%%%%%%%%%%%%%%%%%%%%%%%%%%%%%%%%%%%%

With the detection and the mapping of the 1.809 \MeV\ gamma-ray line 
from \al, gamma-ray line astronomy has made important progress during 
the last 15 years.
\al\ produced in type~II supernova explosion has been proposed by
\citet{arnett77} and \citet{ramaty77} as the most promising gamma-ray 
line emitter, and indeed, only few years later, the {\em HEAO 3} 
experiment discovered the characteristics 1.809 \MeV\ line emission 
from the general direction of the galactic centre at a level of 
$5\sigma$ \citep{mahoney84}.
The intensity of the emission, however, turned out to be in excess of 
the nucleosynthesis calculations at that time, pushing theoreticians 
to look for other \al\ sources.
As a consequence, many candidate \al\ sources, such as novae, red 
giants, Wolf-Rayet stars, or energetic cosmic-ray particle 
interactions have been proposed to fill the gap (see \citet{pd96} for 
a review).
Others suggested more extreme scenarios to resolve the puzzle, such as 
the explosion of a high-metallicity supermassive star near the galactic 
centre \citep{hillebrandt87}, or a nearby recent supernova explosion 
\citep{morfill85}.

These extreme scenarios are now clearly excluded by the first all-sky 
map of the 1.809 \MeV\ gamma-ray line which shows the galactic 
disk as the most prominent emission feature \citep{oberlack96}.
Hence, \al\ production is clearly a galaxywide phenomenon and is not 
dominated by a single and possibly nearby object.
Additionally, the observed intensity profile along the galactic plane 
shows asymmetries and localised emission enhancements, which should be 
characteristic for a massive star population that follows the 
galactic spiral structure \citep{prantzos91}.
Thus, \al\ production seems related to the massive star population.
Correlation studies using tracer maps for various source candidate 
populations strongly support this suggestion.
\citet{knoedl99a} demonstrated that the 1.809 \MeV\ gamma-ray line emission 
follows closely the distribution of galactic free-free emission
which is powered by the ionising radiation of stars with initial 
masses $>20$ \Msol.
This suggests that explosive \al\ production in supernovae may be less 
important than previously thought (e.g.~Timmes et al., 1995), and 
hydrostatic nucleosynthesis in massive mass-losing stars may possibly
be the primary production channel for galactic \al.
Although still a hypothesis, this suggestion is substantiated by the 
correlation of one of the most prominent 1.809 \MeV\ emission features 
in the all-sky map with an agglomeration of young massive star 
associations in the Cygnus region which are lacking recent supernova 
activity (Kn\"odlseder et al., these proceedings).

Having established the correlation between 1.809 \MeV\ emission and 
massive star populations, \al\ becomes an excellent tracer of recent 
galactic star formation.
By refining the knowledge about the 1.809 \MeV\ emission distribution, 
SPI will provide a unique view on the star formation activity in our 
Galaxy, supplementing information obtained at other wavelengths.
To illustrate this potential, the radial \nuc{26}{Al} mass density 
distribution as derived from COMPTEL 1.809 \MeV\ observations is 
shown in Fig. \ref{fig:al26} \citep{knoedl97}.
Apparently, the bulk of galactic star formation occurs at distances of 
less than 6 kpc from the galactic centre.
Star formation is also present within the central 3 kpc of the Galaxy, 
although at a poorly determined rate.
There are indications for enhanced star formation between $3-6$ kpc, 
coinciding with the molecular ring structure as seen in CO data 
\citep{dame87}.
Enhanced star formation is also seen in the solar neighbourhood 
($8-9$ kpc) which probably corresponds to the local spiral arm 
structure.

The radial \nuc{26}{Al} profile is probably not directly 
proportional to the radial star formation profile since \nuc{26}{Al} 
nucleosynthesis may depend on metallicity (e.g.~Meynet, 1994).
It will be important to determine this metallicity dependence in order 
to extract the true star formation profile from gamma-ray line data.
Valuable information about the metallicity dependence will come from a 
precise determination of the 1.809 \MeV\ longitude profile by SPI, and 
the comparison of this profile to other tracers of star formation 
activity, such as galactic free-free emission.
Additionally, observations of gamma-ray lines from \nuc{60}{Fe}, an 
isotope that is mainly believed to be produced during supernova 
explosions (see Section \ref{sec:fe60}), may help to distinguish between 
hydrostatically and explosively produced \nuc{26}{Al}, and therefore 
may allow disentangling the metallicity dependencies for the different 
candidate sources.

A precise determination of the 1.809 \MeV\ latitude profile by SPI
will provide important information about the dynamics and the mixing 
of \al\ ejecta within the interstellar medium.
High velocity \al\ has been suggested by measurements of a broadened 
1.809 \MeV\ line by the {\em GRIS} spectrometer \citep{naya96}, although 
this observation is at some point inconsistent with the earlier 
observation of a narrow line by {\em HEAO 3} \citep{mahoney84}.
In any case, the propagation of \al\ away from its origin should lead 
to a latitude broadening with respect to the scale height of the 
source population, and the observation of this broadening may allow 
the study of galactic outflows and the mass transfer between disk and 
halo of the Galaxy.
Actually, COMPTEL 1.809 \MeV\ observations restrict the scale height 
of the galactic \al\ distribution to $z < 220$ pc 
\citep{knoedl97,oberlack97,diehl97}, which certainly excludes a 
ballistic motion of \al\ at a speed of 500 km s$^{-1}$ (as suggested 
by GRIS), and which could even be in conflict with the reacceleration 
scenario, discussed by Sturner (these proceedings) to explain the 
GRIS observations.
The excellent energy resolution of SPI will easily allow to decide 
whether the 1.809 \MeV\ line is broadened or not, and the improved
angular resolution and sensitivity with respect to COMPTEL will 
allow to determine the scale height of the galactic \al\ distribution 
much more precisely.

The expected energy resolution of SPI of $\sim2.5$ keV at 1.809 \MeV\ 
converts into a velocity resolution of $\sim400$ km s$^{-1}$, allowing 
for line centroid determinations of the order of $50$ km s$^{-1}$ for 
bright emission features.
Thus, in the case of an intrinsically narrow 1.809 \MeV\ line, line 
shifts due to galactic rotation should be measurable by SPI 
\citep{gehrels96}.
Although this objective figures certainly about the most ambitious 
goals of SPI observations, a coarse distance determination of 1.809 \MeV\ 
emission features based on the galactic rotation curve seems in 
principle possible.

Complementary to the study of the large-scale distribution of the 
1.809 \MeV\ emission by SPI will be observations of nearby, 
localised 1.809 \MeV\ emission regions, such as Vela, Cygnus, Carina, 
or Orion.
The aim of these observations will be the identification of emission 
counterparts at other wavelengths in order to associate the 
nucleosynthesis activity to individual objects or specific groups such 
as OB associations or young open clusters.
Already with COMPTEL data, such studies have proven to provide 
important insights in the nature of the \al\ progenitor sources, and 
in constraining nucleosynthetic yields for individual objects
(del Rio et al., 1996; Oberlack et al., 2000, Diehl et al.;
Kn\"odlseder et al.; Lavraud et al.; Pl\"uschke et al., these 
proceedings).

To illustrate the potential of SPI, a contour map of the 1.809 \MeV\ 
emission in the Vela region obtained by COMPTEL is shown in Fig. 
\ref{fig:al26}.
There is a wealth of potential \al\ sources in this field, but the 
limited angular resolution of COMPTEL does not allow for a clear 
identification of the dominant contributors.
Additionally, the sensitivity of COMPTEL is not sufficient to clearly 
separate diffuse from point-like emission \citep{knoedl99b}, leading 
to an additional uncertainty in the association of emission 
structures with \al\ sources.
With improved sensitivity and angular resolution, SPI will help to 
overcome this problem.
Deep exposures of localised emission features will sufficiently 
constrain the 1.809 \MeV\ morphology to associate the structure with 
candidate sources in the field.
In the Vela region, which is part of the {\em INTEGRAL} core program, 
a detection of the Wolf-Rayet star $\gamma$-Vel is awaited, and the 
contributions of the Vela SNR, the new RX J0852.0-4622 supernova 
remnant, and OB star associations should be measurable.
In the Cygnus region, the young globular cluster Cyg OB2 should be 
detectable by SPI, allowing the study of nucleosynthesis in an 
individual massive star association (Kn\"odlseder et al., these 
proceedings).
In Carina, the point-like 1.809 \MeV\ emission feature detected by 
COMPTEL \citep{knoedl96} may be resolved, and the contributions from 
individual young massive star clusters identified.
And in Orion, hints for 1.809 \MeV\ emission from a nearby OB 
association \citep{oberlack97} may be confirmed, enabling the study 
of nucleosynthesis in one of the closest and best studied star 
forming regions in the Milky Way.

%%%%%%%%%%%%%%%%%%%%%%%%%%%%%%%%%%%%%%%%%%%%%%%%%%%%%%%%%%%%%%%%%%%%%%%%%%%%%%%%
\subsection{\fe\ -- supernova nucleosynthesis ?}
\label{sec:fe60}

Although the 1.173 and 1.332 \MeV\ gamma-ray lines from radioactive 
decay of \fe\ produced in core-collapse supernovae were predicted by 
\citep{ramaty77} at an intensity comparable to the 1.809 \MeV\ line, no 
detection of these lines has been reported so far.
The most stringent limits actually come from {\em SMM} \citep{leising94} 
and {\em GRIS} \citep{naya98} observations which constrain the 
$\fe\ / \al$ flux ratio to less than $0.15$, corresponding to less 
than $1.7$ \Msol\ of \fe\ in the present interstellar medium.

Theoretical nucleosynthesis calculations, however, are still quite 
uncertain, and yield determinations for particular isotopes may vary 
by large factors (up to 10) when the input physics are changed 
\citep{woosley99}.
For \al\ and \fe, the biggest uncertainties come from poorly known 
stellar physics, such as convection, mass loss, and rotation, since 
the majority of both isotopes is produced hydrostatically prior to the 
supernova explosion.
Indeed, modern nucleosynthesis calculations for core collapse 
supernovae predict $\fe\ / \al$ flux ratios of only $0.16 \pm 0.10$, 
similar but still compatible to the observed upper limit \citep{timmes95}.
Note that the lower flux ratio (with respect to Ramaty and 
Lingenfelter, 1977) is primarily due to enhanced \al\ production in 
modern models, and not so much due to reduced \fe\ yields.

From a comparison of the abundances of isotopes that are co-produced 
with \fe, \citet{leising94} estimate the galactic \fe\ mass to $0.9$ 
\Msol\ (irrespectively of the production site), which implies a 
$\fe\ / \al$ flux ratio of $0.08$.
At this level, the \fe\ lines should be clearly detectable by SPI.
Traditionally, \fe\ is only considered as a product of core-collapse
supernova nucleosynthesis (e.g. Diehl \& Timmes, 1998), hence the detection 
of the \fe\ lines would provide a unique tracer to study the galactic 
supernova activity and distribution.
The comparison of \fe\ with \al\ observations would then allow to 
separate supernova \al\ production from stellar-wind ejection,
answering the question about the origin of \al.
However, as for \al, other \fe\ nucleosynthesis sites are also in 
discussion.
\citet{woosley97} studied carbon deflagration of type Ia supernovae 
that produce significant amounts of \fe\ in the neutron-rich central 
zones, \citet{gallino99} propose intermediate mass AGB star as a serious 
concurrent for galactic \fe\ production, and \citet{arnould97}
discuss \fe\ production by hydrostatic helium burning in Wolf-Rayet
stars.
Thus, once detected, the \fe\ story may become more complex than 
actually thought.

%%%%%%%%%%%%%%%%%%%%%%%%%%%%%%%%%%%%%%%%%%%%%%%%%%%%%%%%%%%%%%%%%%%%%%%%%%%%%%%%
\subsection{511 keV annihilation radiation}

The 511 keV gamma-ray line due to annihilation of positrons and 
electrons in the interstellar medium has been observed by numerous 
instruments (see \citet{harris98} and references therein).
At least two galactic emission components have been identified so far:
an extended bulge component and a disk component.
Indications of a third component situated above the galactic centre 
have been reported \citep{purcell97,harris98}, yet still needs
confirmation by more sensitive instruments (see also Milne et al., 
these proceedings).

The galactic disk component may be explained by radioactive positron 
emitters, such as \nuc{26}{Al}, \nuc{44}{Sc}, \nuc{56}{Co}, and 
\nuc{22}{Na} \citep{lingenfelter89}.
Although all these isotopes are also gamma-ray line emitters,
only \nuc{26}{Al} will lead to correlated gamma-ray line and 511 keV 
emission since the typical annihilation time scale of some $10^{5}$ 
yrs considerably exceeds the lifetime of the other isotopes.
Consequently, 511 keV line-emission is a potential tracer of extinct 
short-lived galactic radioactivities.

The origin of the galactic bulge component is much less clear.
The observation of an abrupt {\em turn off} of the 511 keV emission 
from the galactic centre by {\em HEAO 3} at the beginning of the 
eighties has led to the idea that a compact object might be responsible 
for the galactic bulge emission \citep{riegler81}.
However, a re-analysis of the same data by \cite{mahoney94} has 
shown that the reported flux variation is not significant.
Also, contemporaneous observations with the {\em SMM} satellite
\citep{share88} and latest observations by OSSE \citep{purcell93} and 
{\em TGRS} \citep{teegarden96} show no evidence for any time-variability.
\citet{lingenfelter89} therefore proposed a two component model for 
the galactic bulge emission which is composed of a variable compact source 
near the galactic centre and a steady diffuse interstellar annihilation 
source.
The flux level of the variable point source, however, is limited by 
the observations to less than $4\,10^{-4}$ \funit, and indeed,
time-variability is not required by the data 
\citep{share90,purcell93,teegarden96}.
On the other hand, the observation of broadened and red-shifted 
annihilation features from 1E~1740.7-2942 \citep{bouchet91} and Nova 
Muscae \citep{goldwurm92} has been considered as evidence for positron 
production in compact objects (but see Section \ref{sec:xnovae}).
However, the contemporaneous observation by OSSE and SIGMA of an outburst 
of 1E~1740.7-2942 in September 1992 gave contradictory results 
\citep{jung95}, casting some doubt on the contribution of compact objects 
to the galactic positron budget.

The origin of the galactic bulge positrons will be one of the 
key-questions addressed by SPI.
Narrow-line transient features with fluxes of $4\,10^{-4}$ \funit\ 
should be detectable by SPI within less than one hour.
If the feature is broadened by 300 keV (as observed for example in 
1E~1740.7-2942) the required observation time $t_{\mbox{broad}}$ 
increases like $t_{\mbox{broad}} = t_{\mbox{narrow}} \Delta E_{\mbox{broad}} / 
\Delta E_{\mbox{GeD}}$, resulting in a detection within $\sim12$
hours.\footnote{The expected narrow line sensitivity of SPI at 511 keV is 
limited by a strong instrumental line to $2\,10^{-5}$ \funit\ for an 
observation time of $10^{6}$ seconds; however, if the line is substantially 
broadened, the sensitivity of $7\,10^{-6}$ \funit\ that is found adjacent 
to 511 keV energies should be applied.}
Hence, the weekly galactic plane scan together with the central radian 
deep exposure performed during the {\em INTEGRAL} core program (Winkler, 
these proceedings) will provide a unique survey of the galactic bulge,
capable of detecting even faint transient 511 keV emission events.

In addition to the detection of transient events, SPI will also provide a 
detailed map of 511 keV emission from the Galaxy.
Using this map, the morphology of the galactic bulge can be studied in 
detail, and the question on the contribution of point sources to the 
galactic bulge emission can be addressed.
In particular, the ratio between bulge and disk emission, which is only 
poorly constrained by existing data (see Milne et al., these 
proceedings), will be measured more precisely, allowing for more 
stringent conclusions about the positron sources of both components.
The 511 keV map will also answer the question about the reality of the
{\em positive latitude enhancement}, which may provide interesting 
clues on the activity near to the galactic nucleus \citep{dermer97}.

Additionally, the 511 keV line shape carries valuable information about 
the annihilation environment which will be explored by SPI.
The dominant annihilation mechanism sensitively depends on the temperature, 
the density, and the ionisation fraction of the medium, and the measurement 
of the 511 keV line width allows the determination of the 
annihilation conditions \citep{guessoum91}.
Observations of a moderately broadened 511 keV line towards the 
galactic centre indicate that annihilation in the bulge mainly occurs
in the warm neutral or ionised interstellar medium \citep{harris98}.
By making spatially resolved line shape measurements, SPI will allow 
to extend such studies to the entire galactic plane, complementing 
our view of galactic annihilation processes.

With its good continuum sensitivity, SPI will also be able to detect 
the galactic positronium continuum emission below 511 keV.
The intensity of this component with respect to that of the 511 keV 
line carries complementary information about the fraction $f$ of 
annihilations via positronium formation, probing the thermodynamic and 
ionisation state of the annihilation environment \citep{guessoum91}.

%%%%%%%%%%%%%%%%%%%%%%%%%%%%%%%%%%%%%%%%%%%%%%%%%%%%%%%%%%%%%%%%%%%%%%%%%%%%%%%%
\subsection{Observation of novae}

Novae explosions are potential sources of various radioactivities, of 
which the nuclei \nuc{7}{Be}, \nuc{13}{N}, \nuc{18}{F}, \nuc{22}{Na}, 
and \nuc{26}{Al} are the most important ones \citep{gomez98a}.
\nuc{7}{Be} decays by electron capture into \nuc{7}{Li}, and the 
observation of the accompanying decay line at 478 keV will be an 
irrefutable prove of cosmic \nuc{7}{Li} enrichment by novae 
\citep{starrfield78}.
Although \nuc{13}{N} and \nuc{18}{F} are no gamma-ray emitters they 
decay under positron emission, and their annihilation in the nova 
envelope may lead to a prominent 511 keV line feature accompanied by 
a low-energy Compton continuum \citep{hernanz97}.
\nuc{22}{Na} maybe produced in the subclass of oxygen-neon (ONe) rich novae, 
and the observation of the 1.275 \MeV\ gamma-ray decay line holds 
valuable information about the dredge-up of white-dwarf core matter 
during the explosion \citep{weiss90}.
Finally, \al\ produced in ONe novae may contribute to the 
galactic \al\ budget, although both theory and observations do not 
suggest a large contribution \citep{kolb97,knoedl99}.

Neither of the nova lines has been detected so far 
\citep{harris91,iyudin95,harris99}, hence any positive detection by 
SPI would be a great piece of information for understanding the nova 
phenomenon and the implied nucleosynthesis processes.
Observationally, novae are situated at the border between diffuse and 
point-like emission sources.
Assuming a galactic nova rate of $35$ yr$^{-1}$ \citep{shafter97} and 
an ONe type proportion of 22\% \citep{livio94}, an average of 
29 such objects should be visible during one lifetime of the 
\nuc{22}{Na} isotope.
For carbon-oxygen (CO) novae, which are the dominant sources of 
\nuc{7}{Be}, an average of 6 objects emitting the 478 keV line should 
be always present in the Galaxy.
Since novae are believed to occur dominantly in the inner Galaxy, the 
limited angular resolution of SPI will recognise them as diffuse emission 
from the central galactic radian \citep{jean00}.
After two years of operations, the core programme exposure of the 
central radian will allow SPI to detect diffuse 478 keV and 1.275 \MeV\ 
fluxes of about $10^{-5}$ \funit, considerably fainter than actual 
upper limits \citep{harris91,iyudin95,leising98}.

The detectibility of individual novae by SPI will, of course, depend 
on the ejected masses of radioisotopes, and the intrinsic width of the 
decay lines (see Eq. \ref{eq:broad}).
Classical CO novae show ejection velocities of $250-2500$ km 
s$^{-1}$, corresponding to Doppler-broadenings of $0.8-8$ keV 
(FWHM) at 478 keV \citep{cohen85}.
ONe novae explode more violently, with ejection velocities 
ranging from $1000-5000$ km s$^{-1}$, leading to line-widths of 
$10-40$ keV (FWHM) at 1.275 \MeV\ \citep{smits91,hayward92}.
However, the ejecta may considerably slow-down during the lifetime 
of \nuc{22}{Na} (e.g. Austin et al., 1996), and the most opportune 
situation for SPI to detect the 1.275 \MeV\ line may occur only 
$1-2$ years after the actual nova explosion.

Estimates of nova nucleosynthetic yields by \citet{hernanz99a} suggest 
that \nuc{7}{Be} decay from individual CO novae within 500 pc should 
be detectable by SPI, \nuc{22}{Na} in ONe novae may be visible out to 
2 kpc, and the 511 keV line from \nuc{13}{N} and \nuc{18}{F} positrons 
could be observed for novae within 3 kpc from the Sun.
Within the quoted distances, the expected event frequencies are
$0.01$, $0.2$, and $0.6$ yr$^{-1}$ for 478 keV, 1.275 \MeV, and 511 
keV line detections, respectively.
Note, however, that the 511 keV emission lasts only for about one day 
\citep{hernanz97}, hence despite the relatively high event frequency, 
a direct detection within the field-of-view of SPI would be extremely 
fortuitous.

However, the short-lasting 511 keV emission with the accompanied 
Compton-continuum from novae could be detected as an increase in the 
counting rate of the SPI BGO collimator and anticoincidence shield 
\citep{jean99}.
In fact, the 91 BGO crystals (making a total detector mass of 512 kg), 
which define the SPI field-of-view and shield the detector array against 
background radiation, provide a formidable large-area detector, that 
will also be inserted in the third interplanetary network to allow 
burst localisations to within $\sim10$ arcmin (Hurley, these 
proceedings).
Estimations based on the revised \nuc{13}{N} and \nuc{18}{F} yield 
predictions by \citet{hernanz99b} suggest that SPI should be able to 
detect massive ONe novae up to distances of $7-8$ kpc, leading to an 
event frequency of $3$ yr$^{-1}$.
However, only limited localisation capabilities are provided by the 
BGO shield, and the search for a counterpart in the visible or the 
near-infrared will be challenging.

%%%%%%%%%%%%%%%%%%%%%%%%%%%%%%%%%%%%%%%%%%%%%%%%%%%%%%%%%%%%%%%%%%%%%%%%%%%%%%%%
\subsection{\nuc{44}{Ti} -- unveiling recent supernova}

Until recently, the census of recent galactic supernova events was 
exclusively based on historic records of optical observations and 
amounted to 6 events during the last 1000 years.
Due to galactic absorption and observational bias, this census is by 
far not complete.
Gamma-ray line observations of the \nuc{44}{Ti} isotope have the 
potential to considerably increase the statistics.
\nuc{44}{Ti} is believed to be exclusively produced by $\alpha$-rich 
freeze-out in supernova events, and the existence of its decay product 
\nuc{44}{Ca} makes this production mandatory.
Due to the penetrating power of gamma-rays, \nuc{44}{Ti} lines from recent 
supernova events throughout the Galaxy can reach the Earth, and 
therefore, unveil yet unknown young supernova remnants.

The prove of principle was achieved by the observation 
of a 1.157 \MeV\ gamma-ray line from the 320 years old Cas A supernova 
remnant using the COMPTEL telescope \citep{iyudin94}.
Evidence for another galactic \nuc{44}{Ti} source was found 
in the Vela region where no young supernova remnant was known before 
\citep{iyudin98}.
Triggered by this discovery, unpublished ROSAT X-ray data showing a 
spherical structure at the position of the new \nuc{44}{Ti} source 
were reconsidered and lead to the discovery of a new supernova 
remnant, now called RX J0852.0-4622 \citep{aschenbach98}.
% a re-analysis of ROSAT X-ray data indeed 
% revealed a spherical structure at the position of the new \nuc{44}{Ti} 
% source, now identified as the RX J0852.0-4622 supernova remnant 
% \citep{aschenbach98}.
In the meanwhile, the remnant has also been discovered at radio 
wavelengths \citep{combi99,duncan00}.
Although the \nuc{44}{Ti} observation is only marginal 
\citep{schoenfelder99}, it is the first time that gamma-ray line 
observations triggered the discovery of a new supernova remnant.
From the X-ray data, an age of less than 1500 yrs and a distance 
$<1$~kpc has been inferred.
Adding the \nuc{44}{Ti} observations further constrains the age and 
distance to $\sim680$ yrs and $\sim200$ pc, respectively 
\citep{aschenbach99}.
Interestingly, nitrate abundance data from an Antarctic ice core 
provide evidence for a nearby galactic supernova $680\pm20$ yrs ago, 
compatible with the \nuc{44}{Ti} data \citep{burgess99}.

Given the marginal detection of the 1.157 \MeV\ line from 
RX J0852.0-4622, a confirmation of \nuc{44}{Ti} decay by SPI will be 
crucial for the further understanding of this object.
Recall, however, that the SPI sensitivity limit depends on the 
intrinsic line width, and substantial Doppler-broadening could 
hamper a convincing detection.
On the other side, the hard X-ray lines at 68 and 78 keV may also be 
detected by IBIS and possibly SPI (Iyudin et al., these proceedings),
and indeed, detecting \nuc{44}{Ti} may be more easily achieved in these 
lines in case of substantial broadening (Georgii et al., these 
proceedings).
Finally, \nuc{44}{Ti} line-profile measurements will provide complementary 
information on the expansion velocity and dynamics of the most inner 
layers of the supernova ejecta.

The regular galactic plane scans and the deep exposure of the central 
radian will lead to a substantial exposure build-up, enabling the detection 
of further hidden young galactic supernova remnants through \nuc{44}{Ti} 
decay.
The observed supernova statistics may then set interesting constraints on 
the galactic supernova rate and the \nuc{44}{Ti} progenitors.
Indeed, actual observations already indicate that some of the galactic 
\nuc{44}{Ca} may have been produced by a rare type of supernova (e.g. 
Helium white dwarf detonations) which produces very large amounts of 
\nuc{44}{Ti} \citep{the99}.

Finally, we want to mention that a detection of the \nuc{44}{Ti} lines
from SN 1987A could be feasibly, and any non-detection will impose 
interesting constrains on $\alpha$-rich freeze-out nucleosynthesis in 
this event \citep{woosley91}.
Fits to the bolometric light curve indicate the presence of 
$10^{-4}$ \Msol\ of \nuc{44}{Ti} in SN 1987A \citep{fransson98}, and 
values twice as high are not excluded by nucleosynthesis theory 
\citep{woosley91} and probably also not by the observations.
Thus, assuming a high \nuc{44}{Ti} yield of $2\,10^{-4}$ \Msol\ and a 
distance of $45$ kpc to the LMC (as suggested by recent 
observations of eclipsing binaries; Guinan et al., 1998; Fitzpatrick 
et al., 2000), a 1.157 \MeV\ line flux of $8\,10^{-6}$ \funit\ is expected, 
of the same order as the SPI sensitivity limit for a narrow line.
Again, line-broadening will worsen the detectibility, but the 
possibility of a detection is certainly worth a deep exposure of the 
LMC by {\em INTEGRAL}.

%%%%%%%%%%%%%%%%%%%%%%%%%%%%%%%%%%%%%%%%%%%%%%%%%%%%%%%%%%%%%%%%%%%%%%%%%%%%%%%%
\subsection{Type Ia supernovae}

The possibility of detecting gamma-ray lines from supernova remnants 
has been first suggested by \citet{clayton65} to identify transbismuth 
radioactivity as tracer of spontaneous \nuc{254}{Cf} fission,
capable to explain the light-curves of type Ia supernovae.
After realising that the subsequent radioactive decays of \nuc{56}{Ni} 
and \nuc{56}{Co} may also explain the light-curves, \citet{clayton69} 
suggested that gamma-ray telescopes should search for their decay lines
from type Ia supernovae within a few Mpc.
Observationally, type Ia events are favoured over the other supernova 
classes because they produce an order of magnitude more radioactive 
\nuc{56}{Ni} than the other types, and they expand rapidly enough to 
allow the gamma-rays to escape before all the fresh radioactivity has 
decayed.
Nevertheless, the first and only direct prove of \nuc{56}{Co} and 
\nuc{57}{Co} radioactivities in a supernova were obtained from 
observations of SN 1987A in the LMC, a type II event.
However, it must be clear that such a nearby event is extremely rare, 
and the chance of observing a type II supernova during the {\em 
INTEGRAL} lifetime is extremely low \citep{timmes97}.

From the SPI sensitivity and the observed type Ia supernova rates 
together with standard models of type Ia nucleosynthesis, one may 
estimate the maximum detectable distance for a type Ia event to about 
$15$ Mpc,\footnote{Again, the expected line-broadening in type Ia supernova 
will reduced the SPI sensitivity with respect to the narrow-line
performances \citep{gomez98b}.}
and the event frequency to one event each 5 years (see also Timmes \& 
Woosley, 1997).
Hence, during an extended mission lifetime of 5 years, SPI has 
statistically spoken the chance to detect one such event.

Nevertheless, {\em CGRO} observations have taught us that supernovae 
are intriguing objects, and even at the detection threshold, their 
observation may provide interesting implications on the progenitor 
nature or explosion mechanism.
For example, \citet{morris95,morris97} report the detection of the 
unusually bright SN 1991T in NGC 4527 (at a distance between $13-17$ 
Mpc) by COMPTEL, implying a \nuc{56}{Ni} mass of $1.3-2.3$ \Msol,
although an earlier analysis of COMPTEL data by \citet{lichti94} and 
a search of \citet{leising95} in OSSE data gave only upper limits.
Assuming that COMPTEL indeed detected SN 1991T, the derived 
\nuc{56}{Ni} mass exceeds all theoretical expectations, requiring 
possibly a super-Chandrasekhar scenario to explain the observations.
Indeed, high ($\sim1$ \Msol) \nuc{56}{Ni} masses have also been inferred
by \citet{spyromilio92} from the strength of forbidden transitions of 
Fe {\scriptsize II} and Fe {\scriptsize III} in late-type spectra, 
by \citet{mazzali95} from modelling early-type spectra with a Monte 
Carlo code, and by \citet{cappellaro97} from fitting 
the long term luminosity evolution by a Monte-Carlo light curve model.
Using premaximum spectra, \citet{ruiz92} inferred a peculiar chemical 
composition structure where a shell of intermediate-mass elements is 
sandwiched between a core and an outer-layer composed of Fe-peak 
elements.
In addition, \citet{fisher99} recognised that the Fe-rich core and 
outer layer are moving faster than the intermediate unburned carbon-rich 
region, a chemico-dynamical configuration that can not be reconciled 
with classical type Ia supernova models.
They suggest that SN 1991T was the result of a white-dwarf merger, 
which exploded by the action of huge tidal forces at the onset of the 
merger due to synchronous rotation \citep{iben97}, which eventually may 
lead to a super-Chandrasekhar \nuc{56}{Ni} mass.
Although this scenario is certainly quite speculative, it illustrates 
that gamma-ray line observations of type Ia supernova may play a 
decisive role in identifying such peculiar events.

Another interesting example is SN 1998bu in M96, which shows the 
characteristics of a rather typical type Ia event at a distance of 
about $11$ Mpc.
Theoretical nucleosynthesis models predict that the radioactive decay 
of \nuc{56}{Co} in SN 1998bu should lead to a peak flux of 
$(1-5)\,10^{-5}$ \funit\ in the 847 keV line, where the upper value 
corresponds to Helium detonation models and the lower value 
represents deflagration models \citep{gomez98b,hoeflich98}.
However, SN 1998bu was observed by OSSE for over 140 days and by COMPTEL 
for almost 90 days without any positive detection \citep{leising99,georgii00}.
The upper time-averaged 847 keV flux limit of OSSE amounts to $3\,10^{-5}$ 
\funit, the COMPTEL limit of $4\,10^{-5}$ \funit is comparable.
For the 1.238 \MeV\ line, COMPTEL imposes an even more stringent flux 
limit of $2\,10^{-5}$ \funit.
Thus, the observations start to constrain type Ia supernova models, 
excluding for example the {\em Helium cap} model for SN 1998bu 
\citep{georgii00}.

Observing SN 1998bu by SPI would probably have been a major 
breakthrough for observational gamma-ray line astrophysics.
Even if the 847 keV line would have been broadened to $50$ keV, which 
is probably rather pessimistic \citep{isern97}, SPI would have achieved 
a sensitivity of $10^{-5}$ \funit\ within a comparable exposure time
($\sim100$ days).
At this level, either SN 1998bu would have been clearly detected or 
the non-detection would have ruled out all existing thermonuclear 
supernova models.
Assuming that SN 1998bu was indeed close to detection (at a 847 keV
flux of say $3\,10^{-5}$ \funit), SPI would have detected the 847 keV
line at a significance of about $10\sigma$, allowing for valuable line 
profile studies.

%%%%%%%%%%%%%%%%%%%%%%%%%%%%%%%%%%%%%%%%%%%%%%%%%%%%%%%%%%%%%%%%%%%%%%%%%%%%%%%%
\subsection{Radioactivities from Accretion Disks around Black Holes}
\label{sec:xnovae}

Another interesting site of nucleosynthesis has been proposed by 
\citet{chakrabarti86} who considers the formation of elements in 
sub-Keplerian thick disks around black holes.
\citet{prantzos99} recently presented exploratory calculations for 
nucleosynthesis of radioactive isotope in thick disk models with 
viscosity parameter $\alpha$ in the $10^{-2}-10^{-4}$ range.
The short timescale of a few $10^{3}$ s and the relative low 
temperatures of a few $10^{8}$ K allow for only limited Hydrogen 
burning to take place through the hot CNO, NeNa and MgAl chains.
In such conditions, radioactive isotopes familiar from e.g. nova 
nucleosynthesis (\nuc{7}{Be}, \nuc{22}{Na}, \al) are produced, along 
with the release of positrons by \nuc{13}{N} and \nuc{15}{O}.
Assuming that $\sim50\%$ of the accretion flow is ejected upon 
arrival in the inner disk (by e.g. radiation pressure or some other 
mechanism), he derived SPI detection distances of 2 kpc, 100 pc, and 1 kpc 
for \nuc{7}{Be}, \nuc{22}{Na}, and \al\ for a $10$ \Msol\ black hole 
that is fed by a massive companion at a rate of $2\,10^{-5}$ \Msol/yr 
for $\sim10^{6}$ yrs.

Indeed, observations of high Li abundances in the secondaries of black 
hole binaries may indicate nucleosynthesis during explosive accretion 
events.
\citet{martin96} suggest that the transient emission line observed by 
SIGMA during the 1991 outburst of Nova Muscae \citep{goldwurm92} may 
be due to de-excitation of \nuc{7}{Li} following the radioactive decay 
of freshly produced \nuc{7}{Be}.
Although it remains difficult to explain the high observed flux in the
478 keV feature \citep{yi97} and the short duration of the event ($\sim13$ 
hours; Goldwurm et al., 1992), the search for signposts of 
nucleosynthesis in accretion disks around black holes presents an 
exciting exploratory field for SPI (see also Isern, these proceedings).

%%%%%%%%%%%%%%%%%%%%%%%%%%%%%%%%%%%%%%%%%%%%%%%%%%%%%%%%%%%%%%%%%%%%%%%%%%%%%%%%
%\section{Other gamma-ray lines}
%\label{sec:interaction}

%%%%%%%%%%%%%%%%%%%%%%%%%%%%%%%%%%%%%%%%%%%%%%%%%%%%%%%%%%%%%%%%%%%%%%%%%%%%%%%%
\section{Interaction lines}

The second fundamental process (after radioactive decay) that leads to 
gamma-ray line emission is nuclear de-excitation following energetic 
particle reactions \citep{ramaty79}.
These gamma-rays exhibit a great wealth of spectral features, ranging 
from very narrow lines to broad features, depending on the 
composition and energy spectrum of the energetic particles, and on 
the composition and physical state of the ambient medium.
Observable gamma-ray line emission is expected from many 
astrophysical sites, including solar flares, the interstellar medium, 
neutron stars and black holes, supernova remnants, and active galactic 
nuclei (see Bykov, these proceedings).

\citet{ramaty79} calculated detailed gamma-ray spectra arising from 
energetic particle interactions with the interstellar medium for 
various abundance compositions.
The most prominent features they predict are 
a broadened line at 4.438 \MeV\ from excited states in \nuc{12}{C} 
and \nuc{11}{B},
a broadened line with a narrow component at 6.129 \MeV\ from excited 
\nuc{16}{O},
and a number of narrow or very narrow lines at lower energies, among 
which is the 847 keV line from excited \nuc{56}{Fe}.
They estimated 4.438 \MeV\ and 6.129 \MeV\ line fluxes from the central 
radian of the Galaxy of the order of $10^{-5}-10^{-4}$ \frad, rather 
optimistic for the potential detection with existing gamma-ray telescopes 
(or with SPI).
However, the normalisation of the spectra was rather arbitrary, and 
indeed, observations by {\em SMM} \citep{harris95} and OSSE \citep{harris96} 
suggest that the line fluxes should be below $10^{-4}$ \frad.

Relying on the observation that the Be/Fe abundance ratio is 
independent of metallicity, \citet{ramaty97} proposed to use the 
current epoch galaxywide Be production rate to normalise the 
gamma-ray line intensities.
With this approach, they predict that the nuclear line emission in 
the energy interval $3-7$ \MeV\ from the central radian of the Galaxy 
should not exceed $2.5\,10^{-5}$ \frad.
From a similar calculation, \citet{tatischeff99} find an upper limit of
$5\,10^{-5}$ \frad\ (see also Tatischeff et al., these proceedings).

Can SPI detect gamma-ray lines at this level?
Following the discussion of \citet{jean00}, the SPI sensitivity to 
diffuse emission may be estimated by multiplying the point-source 
sensitivity by $\sqrt{2}$.
Assuming that the galactic plane emission has a latitude extent 
of less than $\pm8\deg$ (i.e. it fits entirely in the SPI fully-coded 
field of view), SPI is able to observe $0.28$ radians of the galactic 
ridge within the fully coded field-of-view of $16\deg$ with a 
single pointing.
Thus, the SPI sensitivity in units of \frad\ for diffuse galactic plane 
emission is obtained by multiplying the point-source sensitivity by a 
factor of 5.
For narrow lines in the $3-7$ \MeV\ range, a sensitivity of $2.5\,10^{-5}$ 
\frad\ seems feasible.

If the excited nuclei are in the gas phase of the interstellar medium, 
the recoil following excitation will lead to a considerable 
Doppler-broadening of typically $\sim100$ keV (FWHM).
Following Eq. \ref{eq:broad} it seems unrealistic that such lines are 
detectable by SPI.
If, however, a large fraction of nuclear de-excitations appear within 
interstellar grains, the nuclei may lose their recoil energy before 
emitting gamma-rays, leading to much narrower lines \citep{lingenfelter77}.
The best candidate for such narrow line emission is the refractory 
isotope \nuc{56}{Fe} (with a line at 847 keV) which is expected to show 
a line width inferior to the spectral resolution of SPI (Tatischeff, 
private communication).
However, flux expectations of the order of $10^{-6}$ \frad\ would 
make the detection of these line a great surprise.

Another site of nuclear excitation may be superbubbles where shock 
acceleration of metal-rich stellar ejecta could lead to gamma-ray line 
emission.
Of course, this scenario was primarily proposed to explain the 
observation of excitation lines in the Orion region \citep{bloemen94}, 
which now after the withdrawal of the results \citep{bloemen99} seems 
less promising.
Nevertheless, gamma-ray line fluxes of $10^{-5}$ \funit\ from nearby 
superbubbles, such as the Orion-Eridanus superbubble, seem possible 
\citep{parizot99}, hence the search for gamma-ray line emission from
such objects is still an interesting goal for SPI.

%%%%%%%%%%%%%%%%%%%%%%%%%%%%%%%%%%%%%%%%%%%%%%%%%%%%%%%%%%%%%%%%%%%%%%%%%%%%%%%%
\section{Neutron capture signatures}

Nuclear reactions within the accretion flow of compact objects may 
not only produce radioisotopes (see Section \ref{sec:xnovae}), but 
can also release neutrons that may subsequently be captured by 
protons to form \nuc{2}{H} (Deuterium).
The neutron capture is accompanied by emission of a 2.223 \MeV\ 
photon that eventually may be observable by gamma-ray telescopes
\citep{brecher80}.
The detection of the 2.223 \MeV\ line would provide evidence for this 
nuclear reactions, and would allow for a detailed study of the 
heating processes within the accretion flow.
Note, however, that the line may eventually be broadened or shifted 
\citep{brecher80,aharonian84} and emission at 2.223 \MeV\ is only 
expected if the neutrons can encounter a medium that is denser and 
cooler than the hot accretion disk plasma \citep{guessoum89}.

With the aim of testing this hypothesis, \citet{mcconnell98} 
assembled an all-sky map for 2.223 \MeV\ line emission from 5 years 
of COMPTEL data.
However, instead of detecting signatures from X-ray binaries within the 
galactic plane, they found only one significant ($3.7\sigma$) point-like 
source located at $(l,b)=(300.5\deg, -29.6\deg)$ with a 2.223 \MeV\ flux 
of $1.7\,10^{-4}$ \funit.
Among the 11 objects catalogued by the {\em ROSAT} all-sky survey that 
coincide with the $3\sigma$ localisation error box of this source, 
none corresponds to any known X-ray binary, cataclysmic variable, or 
active galactic nucleus.
The most interesting of the {\em ROSAT} sources is RE J0317-853, which 
is identified as a nearby ($35$ pc) peculiar white dwarf.
RE J0317-853 is one of the hottest known white dwarfs ($50\,000$ K) 
and shows a magnetic field between 180-800 MG, one of the strongest known 
for white dwarfs.
Regular white light variations indicate an anomalous high rotation period 
of 12 minutes, which is difficult to explain for an isolated white dwarf
(indicating either an unidentified binary companion or a white dwarf 
merger).

\citet{mcconnell98} suggest flaring activity on RE J0317-853 may lead 
to 2.223 \MeV\ line emission, similar to the processes observed in solar 
flares.
Indeed, a strong, highly polarised radio flare has been observed from 
this object in 1996, but more recent radio observations did not show 
evidence for any flaring activity \citep{barrett99}.
Also optical spectroscopy showed no flaring behaviour or evidence for a 
low mass companion, hence \citet{barrett99} conclude that RE J0317-853
is not a strong candidate for the 2.223 \MeV\ counterpart.

In any case, the 2.223 \MeV\ emission feature remains intriguing, and 
SPI observations should shed new light on the nature of this object.
SPI should easily allow to detect the 2.223 \MeV\ line, even if it 
would be broadened to $1$ \MeV\ (FWHM) -- which however is excluded 
by the COMPTEL observations.
Even IBIS should detect the 2.223 \MeV\ source if the flux is as high as 
reported by \citet{mcconnell98}, and the good angular resolution 
should allow to improve the counterpart identification considerably.

%%%%%%%%%%%%%%%%%%%%%%%%%%%%%%%%%%%%%%%%%%%%%%%%%%%%%%%%%%%%%%%%%%%%%%%%%%%%%%%%
\section{Conclusions}

With this review, we tried to illustrate that SPI has been designed 
to address an extremely wide range of research topics, reaching from 
massive star nucleosynthesis over supernova physics to accretion 
flows in compact objects.
Since SPI is the spectrometer on {\em INTEGRAL} we deliberately 
excluded topics related to continuum emission -- those will be 
addressed by the corresponding paper of the IBIS collaboration 
(Ubertini, these proceedings).
Nevertheless, topics like
the diffuse gamma-ray background (Lichti et al., these proceedings),
the diffuse galactic gamma-ray emission (Valinia et al., these 
proceedings),
pulsars (Dyks et al., these proceedings),
active galactic nuclei (Collmar, these proceedings), or
gamma-ray bursts (Mereghetti et al., these proceedings),
are equally important for SPI, which has a continuum sensitivity 
comparable to IBIS, but which is optimised for large-scale emission
(IBIS provides a much better angular resolution but is less sensitive 
to extended emission features).
On the other hand, IBIS may also detect gamma-ray lines and can help 
to identify counterparts by means of the high localisation accuracy.
Thus, SPI and IBIS are complementary instruments on {\em 
INTEGRAL}, which, when combined, will explore the gamma-ray sky 
far beyond the established horizon.

%%%%%%%%%%%%%%%%%%%%%%%%%%%%%%%%%%%%%%%%%%%%%%%%%%%%%%%%%%%%%%%%%%%%%%%%%%%%%%%%
\section*{Acknowledgments}

The authors want to thank
M. Hernanz,
P. Jean,
G. Meynet,
N. Mowlavi,
N. Prantzos, and
V. Tatischeff 
for helpful discussions.

%%%%%%%%%%%%%%%%%%%%%%%%%%%%%%%%%%%%%%%%%%%%%%%%%%%%%%%%%%%%%%%%%%%%%%%%%%%%%%%%
% References
%%%%%%%%%%%%%%%%%%%%%%%%%%%%%%%%%%%%%%%%%%%%%%%%%%%%%%%%%%%%%%%%%%%%%%%%%%%%%%%%

\end{document}